\documentclass[aps,prd,
twocolumn, 
preprintnumbers,groupedaddress]{revtex4}
\usepackage{epsbox}
\usepackage{amsmath}
\usepackage[dvips]{graphicx}
\usepackage{bm}
\begin{document}
\preprint{\vtop{
{\hbox{YITP-09-25}\vskip-0pt
                 \hbox{KANAZAWA-09-05} \vskip-0pt
}
}
}


\title{$\bm{\omega}$-$\bm{\rho^0}$ mixing 
as a possible origin of the hypothetical \\    
isospin non-conservation 
in the $\bm{X(3872)\rightarrow\pi^+\pi^-J/\psi}$ 
decay
}

\author{
Kunihiko Terasaki   
}
\affiliation{\hspace{-3.5mm}
Yukawa Institute for Theoretical Physics, Kyoto University,
Kyoto 606-8502, Japan \\
Institute for Theoretical Physics, Kanazawa University, 
Kanazawa 920-1192, Japan
}

\begin{abstract}
{The ratio of branching fractions 
${Br(X(3872)\rightarrow \gamma J/\psi)}/
{Br(X(3872)\rightarrow \pi^+\pi^-J/\psi)}$
is studied by assuming that the $X(3872)\rightarrow \pi^+\pi^-J/\psi$ 
decay proceeds through the $\rho^0$ meson pole which is caused by the 
$\omega$-$\rho^0$ mixing. 
As the result, it is seen that the calculated ratio is compatible with 
the measured values of the ratio when $X(3872)$ is an axial-vector 
tetra-quark state, while it would be much larger than the measurements 
when it is a charmonium. 
Therefore, the existing data on the ratio seem to favor the 
tetra-quark interpretation of $X(3872)$ over the charmonium, 
although a small mixing of the charmonium is not excluded. 
}
\end{abstract}

\maketitle

A narrow hidden-charm resonance, $X(3872)$, has been observed in the 
$\pi^+\pi^-J/\psi$ mass distribution from the 
$B^+\rightarrow K^+\pi^+\pi^-J/\psi$ decay~\cite{Belle-X-rho}, and its 
mass and width are now compiled as 
$m=3872.2 \pm 0.8$ MeV and $\Gamma=3.0^{+2.1}_{-1.7}$ MeV~\cite{PDG08}. 
In addition, another resonance peak in the $\pi^+\pi^-\pi^0J/\psi$ mass 
distribution has been observed at the same mass. By identifying the 
above two resonances, the ratio of the measured branching fractions has 
been provided as~\cite{Belle-X-omega} 
\begin{equation}
\frac{Br(X(3872)\rightarrow \pi^+\pi^-\pi^0J/\psi)}
{Br(X(3872)\rightarrow \pi^+\pi^-J/\psi)}
=1.0\pm 0.4\pm 0.3.                           \label{eq:3pi/2pi}
\end{equation}
If the above identification 
is the case, 
Eq.~(\ref{eq:3pi/2pi}) would imply that the conservation of $G$-parity 
in the above decays is badly violated 
in contrast with the ordinary strong interactions. 
Besides, it has been noted~\cite{Belle-X-rho,CDF-pipi} 
that the $X(3872)\rightarrow\pi^+\pi^-J/\psi$ decay 
proceeds through the $X(3872)\rightarrow\rho^0J/\psi$. 
If it is the case and the isospin conservation works well in this decay, 
there should exist charged partners of $X(3872)$. 
However, a search for them has given a negative 
result~\cite{Babar-charged-partner}. 
This implies that $X(3872)$ is an iso-singlet state, and hence the
isospin conservation does not work in the  
$X(3872)\rightarrow\rho^0J/\psi\rightarrow\pi^+\pi^-J/\psi$ decay. 
It also has been suggested~\cite{Belle-X-omega} that the 
$X(3872)\rightarrow\pi^+\pi^-\pi^0J/\psi$ decay proceeds through the 
sub-threshold decay $X(3872)\rightarrow \omega J/\psi$. 
If isospin is conserved in this decay, $X(3872)$ would be an iso-singlet 
state. 
This is consistent with the fact that no charged partner of $X(3872)$ 
has been observed. 
Under the above conditions, a phenomenological analysis has provided the 
following ratio of amplitudes~\cite{Suzuki} from Eq.~(\ref{eq:3pi/2pi}), 
\begin{eqnarray}
&&\hspace{-5mm} 
\Bigl|\frac{A(\rho^0\psi)}{A(\omega\psi)}\Bigr|
\equiv
\Bigl|\frac{A(X(3872)\rightarrow \rho^0J/\psi)}
                        {A(X(3872)\rightarrow \omega J/\psi)}\Bigr|
\nonumber\\
&&\hspace{10.5mm}
=0.27\pm 0.02.                                   \label{eq:rho/omega}
\end{eqnarray}
This shows explicitly a large violation of isospin conservation in 
decays of $X(3872)$. 

Not only the above hadronic decays but also the radiative 
$X(3872)\rightarrow \gamma J/\psi$ decay has been observed and the ratio 
of its branching fractions 
\begin{equation}
R \equiv \frac{Br(X(3872)\rightarrow \gamma J/\psi)}
     {Br(X(3872)\rightarrow \pi^+\pi^-J/\psi)}
                                      \label{eq:radiative-fraction-def}
\end{equation}
has been given as 
\begin{equation}
R_{\rm Belle} = 0.14 \pm 0.05 \hspace{2mm}{\rm and}\hspace{2mm}
R_{\rm Babar} = 0.33 \pm 0.12 
                                    \label{eq:radiative-fraction-exp}
\end{equation}
by the Belle collaboration~\cite{Belle-X-omega}, and recently 
by the Babar~\cite{Babar-rad}, respectively. 
From this, the charge conjugation parity of $X(3872)$ would be even, 
if it is tacitly assumed that $X(3872)$ is a single meson state. 
Regarding with its spin-parity, the angular analysis in its decay 
products favors $J^P=1^+$ over the other quantum 
numbers~\cite{Belle-X-J^P}. 

Productions of $X(3872)$ satisfy well the isospin 
symmetry~\cite{production}, 
\begin{equation}
\frac{ Br(B^0\rightarrow X(3872)K^0)} 
            {Br(B^+\rightarrow X(3872)K^+)}
= 0.94 \pm 0.24 \pm 0.10,                            
                                             \label{eq:production}
\end{equation}
in contrast with the $X(3872)\rightarrow\pi^+\pi^-J/\psi$ decay, 

To solve the above puzzle concerning with the $G$-parity 
non-conservation, there might be some possible options. 
One of them is to suppose that $X(3872)$ consists of two (approximately) 
degenerate states with opposite $G$-parities, i.e., to suppose that 
there exist two (approximately) degenerate axial-vector states with 
opposite $G$-parities. 
Such a situation can be realized in a unitarized chiral 
model~\cite{Oset} and also in a tetra-quark model~\cite{Terasaki-X}. 
The second option is to introduce an explicit violation of isospin 
conservation in the $X(3872)$ physics. 
There are some models in this category; for example, a 
molecular model~\cite{molecule}, a diquark-antidiquark 
model~\cite{Maiani}, etc. 
In the molecular model, it is supposed that $X(3872)$ consists 
dominantly of $D^0\bar D^{*0}$ + its charge conjugate state ($c.c.$). 
However, this model cannot reproduce~\cite{Braaten} the isospin symmetry 
in its productions, Eq.~(\ref{eq:production}). 
In the diquark-antidiquark model which is quite different from the 
tetra-quark model in the first option~\cite{Terasaki-X}, 
it has been predicted that two axial-vector $[cd][\bar c\bar d]$ 
and $[cu][\bar c\bar u]$ states exist as approximately independent mass 
eigenstates with a mass difference (at least) $\Delta m = 7\pm 2$ MeV, 
and that their charged partners should exist. 
However, these results are in contradiction to the measured mass 
difference 
$\Delta m_{\rm exp} = 0.22\pm 0.90\pm 0.27$ MeV~\cite{production} 
and the negative result from the search for these charged partners 
as discussed before. 
The above discussions imply that the existing models in the second 
option seem to be unlikely. 
The third option is to consider a dynamical breaking of isospin 
symmetry.  
The intermediate (or final) states in near-threshold decays, for 
example, $D^0\bar D^{*0}$ and $D^+D^{*-}$ states in decays of 
$X(3872)$, can violate the isospin symmetry because of the mass 
differences between $D^0\,\,(D^{*0})$ and 
$D^+\,\,(D^{*+})$~\cite{Suzuki,dynamical}. 
However, it seems to be not yet conclusive if the dynamical breaking 
of isospin symmetry can lead to the measured $\rho^0$ pole dominance 
in the $X(3872)\rightarrow\pi^+\pi^-J/\psi$ decay. 

Therefore, we here propose a new idea to check if the $\rho^0$ pole 
dominance works in the $X(3872)\rightarrow\pi^+\pi^-J/\psi$ decay as 
noted by the Belle and CDF collaborations. 
To this aim, we assume that the isospin non-conservation in decays of 
$X(3872)$ is caused by the phenomenologically known $\omega$-$\rho^0$ 
mixing~\cite{omega-rho} and study if the assumption can be reconciled 
with the ratio of decay rates in Eq.~(\ref{eq:radiative-fraction-exp}) 
and the isospin symmetry in the productions of $X(3872)$. 
First of all, we point out that we do not need to worry about isospin 
symmetry breaking in the productions of $X(3872)$ under this 
assumption, because the $\omega$-$\rho^0$ mixing does not play any 
important role in these processes. 
Under the same assumption, the isospin non-conserving 
$X(3872)\rightarrow \rho^0J/\psi$ decay proceeds through two steps; 
the isospin conserving sub-threshold decay, 
\begin{equation}
X(3872)\rightarrow \omega J/\psi,                 \label{eq:I-conserv}
\end{equation}
and the subsequent $\omega$-$\rho^0$ mixing, 
\begin{equation}
X(3872)\rightarrow\omega J/\psi\rightarrow\rho^0J/\psi.  
                                                   \label{eq:I-viol}
\end{equation}
When the above assumption is combined with the vector meson dominance 
hypothesis (VMD)~\cite{VMD}, the $X(3872)\rightarrow \gamma J/\psi$
decay can proceed through the channels, 
\begin{eqnarray}
&& X(3872)\rightarrow\omega J/\psi\rightarrow\gamma J/\psi
\,\,\,\,{\rm and}\,\,\,\, 
\nonumber\\
&&
X(3872)\rightarrow\omega J/\psi\rightarrow\rho^0 J/\psi \rightarrow 
\gamma J/\psi.                       
                                                \label{eq:X-gamma}
\end{eqnarray}
However, if $X(3872)$ is an axial-vector charmonium, the decay could 
have an extra contribution through the $J/\psi$ pole,
\begin{equation}
X(3872)\rightarrow J/\psi J/\psi\rightarrow\gamma J/\psi,
                                               \label{eq:psi-pole}
\end{equation}
while, if $X(3872)$ is a tetra-quark $\{cn\bar c\bar n\}$ state 
(a tetra-quark meson like $[cn](\bar c\bar n) + (cn)[\bar c\bar n]$ or 
a molecule of dominantly $D\bar D^{*} + c.c.$ in the first option 
mentioned above), such a contribution would be suppressed because of the 
OZI rule~\cite{OZI}. 
In this way, we study if the above isospin non-conservation can be 
reconciled with the measured ratio of branching fractions in 
Eq.~(\ref{eq:radiative-fraction-exp}), and, as the result, we shall see 
that the existing measurements of the ratio seem to favor a certain kind 
of tetra-quark interpretations of $X(3872)$. 

The rate for the $X\rightarrow \gamma\psi$ decay is given by 
\begin{eqnarray}
&&\hspace{-5mm}\Gamma(X\rightarrow \gamma\psi)               
= \frac{1}{3}\Bigl(\frac{q_\gamma}{8\pi m_X^2}\Bigr)
\sum_{\rm spins}
|M(X\rightarrow \gamma\psi)|^2,                                
                                                      \label{eq:rate}
\end{eqnarray}
where $X$ and $\psi$ denote $X(3872)$ and $J/\psi$, respectively, 
(we use this notation hereafter), and $q_\gamma$ is the center-of-mass 
momentum of $\gamma$. 
In Eq.~(\ref{eq:rate}), the amplitude $M(X\rightarrow \gamma\psi)$ can 
be written as  
\begin{eqnarray}
&& \hspace{-8mm} M(X(P)\rightarrow \gamma(k)\psi(p)) 
\nonumber\\    &&\hspace{-8mm} 
= A(X(P)\rightarrow \gamma(k)\psi(p))
\tilde M(X(P)\rightarrow \gamma(k)\psi(p)), 
                                       \label{amp-X-psi-gamma-general}
\end{eqnarray}
where $P$, $p$ and $k$ are the momenta of $X$, $\psi$ and $\gamma$, 
respectively. 
The kinematical factor $\tilde M(X(P)\rightarrow \gamma(k)\psi(p))$ can 
be provided in the form 
\begin{eqnarray}
&&\hspace{-10mm} \tilde M(X(P)\rightarrow \gamma(k)\psi(p))     
\nonumber\\
&&
= \epsilon^{\mu\nu\alpha\beta}
e_\mu(X;P)e_\nu(\psi;p)F_{\alpha\beta}(\gamma;k), 
                                             \label{kinematical-gamma}
\end{eqnarray}
by the polarization vectors $e_\mu(X;P),\,\,e_\nu(\psi;p)$ and 
$F_{\alpha\beta}(\gamma;k)
=\frac{1}{2}[k_\alpha e_\beta(\gamma;k) - k_\beta e_\alpha(\gamma;k)]$,    
and the truncated amplitudes for the radiative decays in 
Eq.~(\ref{eq:X-gamma}) by 
\begin{equation}
A(X\rightarrow\omega\psi \rightarrow\gamma\psi)_\omega 
\simeq g_{X\omega\psi}\frac{X_\omega(0)}{m_\omega^2}, 
                                           \label{eq:radiative-omega}
\end{equation}
and 
\begin{equation}
A(X\rightarrow\omega\psi\rightarrow\rho^0\psi
\rightarrow\gamma\psi)_{\rho^0}
\simeq g_{X\omega\psi}\frac{g_{\omega\rho}}{m_\omega^2}
\frac{X_\rho(0)}{m_\rho^2}, 
                                           \label{eq:radiative-rho}
\end{equation}
respectively. Here, it has been assumed that $g_{\omega\rho}$ is not 
very sensitive to $k^2$ in the region under consideration. 
It might be understood by analogy with the $\gamma$-$\rho^0$ and 
$\gamma$-$\omega$ coupling strengths, i.e., $X_\rho(0)$ and 
$X_\omega(0)$ on the photon-mass-shell are close to $X_\rho(m_\rho^2)$ 
and $X_\omega(m_\omega^2)$ on the $\rho$ and $\omega$ mass-shells, 
respectively~\cite{VMD-Terasaki}. 
In this case, the size of the amplitude in Eq.~(\ref{eq:radiative-rho}) 
is much smaller than that in Eq.~(\ref{eq:radiative-omega}), because 
$|{g_{\omega\rho}}/{m_\omega^2}|\ll 1$ as seen later, so that the 
$\rho^0$ pole contribution can be neglected. 
The extra contribution through the $\psi$ pole is written as  
\begin{equation}
A(X\rightarrow\psi\psi \rightarrow\gamma\psi)_\psi 
\simeq 2g_{X\psi\psi}\frac{X_\psi(0)}{m_\psi^2}. 
                                           \label{eq:radiative-psi}
\end{equation}
Therefore, the total truncated amplitude for the radiative decay is 
given by 
\begin{eqnarray}
&&\hspace{-8mm} A(X\rightarrow \gamma J/\psi)              
\simeq g_{X\omega\psi}\frac{X_\omega(0)}{m_\omega^2}
\bigl\{1 + K \bigr\},  
                                         \label{eq:total-amp-rad}
\end{eqnarray}
where $K$ is defined by 
\begin{equation}
K = 2\frac{m_\omega^2}{m_\psi^2}\cdot \frac{X_\psi(0)}{X_{\omega}(0)}
\cdot \frac{g_X\psi\psi}{g_{X\omega\psi}}.                \label{eq:K}
\end{equation}
Here, the ${X\psi\psi}$ vertex is OZI-allowed if $X$ is a charmonium, 
while it is suppressed if $X$ is a $\{cn\bar c\bar n\}$ state. 
Because the values of $X_\omega(0)$ and $X_\psi(0)$ have been 
estimated~\cite{VMD-Terasaki} to be  
$X_\omega(0) = 0.011\pm 0.001$ GeV$^2$ and $X_\psi(0)= 0.050\pm 0.013$ 
GeV$^2$, the remaining unknown parameters which are included in the 
calculated rate for the radiative decay 
\begin{eqnarray}
&&\Gamma(X\rightarrow\gamma\psi) \nonumber\\
&&= \frac{q_\gamma}{48\pi m_X^2}
\Bigl\{ \frac{(m_X^2 - m_\psi^2)^2}{m_\psi^2}
+ \frac{(m_X^2 - m_\psi^2)^2}{m_X^2}\Bigr\}   \nonumber\\
&&\hspace{15mm}
\times
\Bigl|g_{X\omega\psi}\frac{X_\omega(0)}{m_\omega^2}
\bigl\{1 + K \bigr\}\Bigr|^2,
                                                  \label{eq:rate-rad}
\end{eqnarray}
are $g_{X\omega\psi}$ and $K$. 
Because of the OZI-rule, 
$|{g_X\psi\psi}/{g_{X\omega\psi}}|\ll 1$ if $X$ is a 
$\{cn\bar c\bar n\}$ state, while 
$|{g_X\psi\psi}/{g_{X\omega\psi}}|\gg 1$ if $X$ is a charmonium, so that 
$|K|$ would be much smaller than unity in the former case, while it 
would be much larger than unity in the latter case, as discussed above. 

The amplitude for the isospin non-conserving 
decay through the $\omega$-$\rho^0$ mixing is written in the form,  
\begin{eqnarray}\hspace{-3mm}
&& M(X\rightarrow\omega\psi\rightarrow\rho^0\psi
\rightarrow\pi^+\pi^-\psi)  \nonumber\\
&&\hspace{3mm} = A(X\rightarrow\omega\psi\rightarrow\rho^0\psi
\rightarrow\pi^+\pi^-\psi)                          \nonumber\\
&&\hspace{7mm}
\times \tilde M(X\rightarrow\omega\psi\rightarrow\rho^0\psi
\rightarrow\pi^+\pi^-\psi),
                                               \label{eq:amp-rhopsi}
\end{eqnarray}
where 
\begin{eqnarray}
&& \hspace{-5mm}
\tilde M(X\rightarrow\omega\psi\rightarrow\rho^0\psi
\rightarrow\pi^+\pi^-\psi)                     \nonumber\\ 
 && \hspace{3mm}
= \epsilon^{\mu\nu\rho\sigma}e_\mu(X;P)e_\nu(\psi;p)k_\rho q_\sigma, 
\nonumber\\
&& \hspace{-5mm}
A(X\rightarrow\omega\psi\rightarrow\rho^0\psi
\rightarrow\pi^+\pi^-\psi)
\nonumber\\
&&\hspace{3mm}
=  g_{X\omega\psi}\Bigl(\frac{g_{\omega\rho}}{m_\omega^2 - k^2}\Bigr)
\Bigl(\frac{g_{\rho^0\pi^+\pi^-}}{m_\rho^2 - k^2}\Bigr).           
\end{eqnarray}
Here, $k = p_{\pi^+} + p_{\pi^-}$ and $q = p_{\pi^+} - p_{\pi^-}$ with 
the momenta $p_{\pi^+}$ and $p_{\pi^-}$ of $\pi^+$ and $\pi^-$, 
respectively. 
Because $\rho^0$ and $\omega$ are resonant states, in particular, the 
former is very broad, we use the Breit-Wigner form~\cite{Pilkuhn} for 
their propagators. 
In this way, we obtain 
\begin{eqnarray}
&& \Gamma(X\rightarrow\omega\psi\rightarrow\rho^0\psi
\rightarrow\pi^+\pi^-\psi) \nonumber\\
&& = \frac{|g_{X\omega\psi}|^2|g_{\omega\rho}|^2
|g_{\rho^0\pi^+\pi^-}|^2}
{2304\pi^3}\Bigl(\frac{m_X^2 + m_\psi^2}{m_X^5m_\psi^2}\Bigr) 
\nonumber\\
&& 
\times\int_{s_{min}}^{s_{max}}ds\Biggl\{
\Biggl[\sqrt{\frac{s - 4m_\pi^2}{s}}
\nonumber\\
&&\hspace{0mm}\times
\frac{(s - 4m_\pi^2)
\sqrt{s^2 - 2(m_X^2 + m_\psi^2)s + (m_X^2 - m_\psi^2)^2}}
{\{(m_\omega^2 - s)^2 + (m_\omega\Gamma_\omega)^2\}
\{(m_\rho^2 - s)^2 + (m_\rho\Gamma_\rho)^2\}}\Biggr]  \nonumber\\
&&\hspace{0mm}\times\Biggl[
s^2 - \frac{2\{(m_X^2 - m_\psi^2)^2 - 2(m_Xm_\psi)^2 \}}
{m_X^2 + m_\psi^2}s \nonumber\\
&&\hspace{40mm}+ (m_X^2 - m\psi^2)^2\Biggr]\Biggr\},     
                                           \label{eq:rate-pipipsi}
\end{eqnarray}
where $s = k^2$, $s_{min}=4m_\pi^2$ and $s_{max}=(m_X - m_\psi)^2$. 
The $\rho^0\pi^+\pi^-$ coupling constant and the $\omega$-$\rho^0$ 
mixing parameter can be estimated to be 
$|g_{\rho^0\pi^+\pi^-}| \simeq 5.98$ from the measured rate~\cite{PDG08} 
$\Gamma(\rho\rightarrow\pi\pi)_{\rm exp} \simeq 149.4$ MeV, and 
$|g_{\omega\rho}|= (3.4\pm 0.5)\times 10^{-3}$ GeV$^2$ from the rate for 
the isospin non-conserving $\omega\rightarrow\pi^+\pi^-$ decay, by 
assuming that it proceeds through the $\rho^0$ pole which is caused by 
the $\omega$-$\rho^0$ mixing and by using the experimental 
data~\cite{PDG08} on the branching fraction 
$Br(\omega\rightarrow\pi^+\pi^-)= 1.53^{+0.11}_{-0.13}$ \% 
and the total width $\Gamma_\omega = 8.49 \pm 0.08$ MeV, where the broad 
width $\Gamma_\rho \simeq 149.4$ MeV has been taken into account. 
(The above value of $|g_{\omega\rho}|$ implies that 
$|g_{\omega\rho}/m_{\omega}^2| \ll 1$ as mentioned before.) 
Therefore, the remaining unknown parameter which is involved in 
$\Gamma(X\rightarrow\omega\psi\rightarrow\rho^0\psi
\rightarrow\pi^+\pi^-\psi)$ is $g_{X\omega\psi}$. 

The calculated rate for the $X\rightarrow\gamma\psi$ decay includes two 
parameters, $g_{X\omega\psi}$ and $K$, as seen in 
Eq.~(\ref{eq:rate-rad}). 
However, when $X$ is a tetra-quark $\{cn\bar c\bar n\}$ state, $K$ 
has been negligibly small, as discussed before. 
Therefore, in this case, the ratio of the rates 
\begin{equation}
R_{\rm tetra}=\frac{\Gamma(X\rightarrow\gamma\psi)}
{\Gamma(X\rightarrow\omega\psi\rightarrow\rho^0\psi
\rightarrow\pi^+\pi^-\psi)}\Bigr|_{\rm tetra}
\end{equation}
contains no unknown parameter, and is estimated to be 
$R_{\rm tetra}\simeq 0.33$ by inserting the central values~\cite{PDG08} 
of the measured masses and widths involved; 
$m_X = 3872.2$ MeV, $m_\psi = 3096.916$ MeV, $m_\rho = 775.49$ MeV, 
$\Gamma_\rho = 149.4$ MeV, $m_\omega = 782.65$ MeV, 
$\Gamma_\omega = 8.49$ MeV, and $m_\pi = 139.57$ MeV. 
This result ($R_{\rm tetra}\simeq 0.33$) seems to be a little bit larger 
than $R_{\rm Belle}$ but is consistent with $R_{\rm Babar}$ in 
Eq.~(\ref{eq:radiative-fraction-exp}). 
In contrast, when $X$ is an axial-vector charmonium, the $X\psi\psi$ 
vertex is OZI-allowed while the $X\omega\psi$ one is suppressed. 
Therefore, the size of $K$ given in Eq.~(\ref{eq:K}) should be much 
larger than unity, and hence the calculated ratio $R_{\{c\bar c\}}$ 
also should be much larger than $R_{\rm tetra} 
(\simeq R_{\rm Babar} \gtrsim R_{\rm Belle} )$. 
This implies that it is difficult to reconcile the $\rho^0$ pole 
dominance in the $X\rightarrow\pi^+\pi^-\psi$ decay with the observed 
ratios, $R_{\rm Belle}$ and $R_{\rm Babar}$, when $X$ is a charmonium. 

In summary, we have studied the ratio of rates for the $\gamma\psi$ 
decay to the $\pi^+\pi^-\psi$ of $X(3872)$, assuming that the isospin 
non-conservation in the $X\rightarrow\pi^+\pi^-\psi$ decay is caused by 
the $\omega$-$\rho^0$ mixing, and we have seen that the measured ratios 
of the decay rates seem to favor a $\{cn\bar c\bar n\}$ interpretation 
of $X(3872)$ over a charmonium (although a small mixing of the latter 
is not excluded), and that the $\rho^0$ pole dominance in the 
$X(3872)\rightarrow\pi^+\pi^-\psi$ decay might be understood by the 
$\omega$-$\rho^0$ mixing in consistency with the isospin non-conserving 
$\omega\rightarrow\pi^+\pi^-$ decay, when $X$ is a $\{cn\bar c\bar n\}$ 
state. 

Because the existing data on the decays of $X(3872)$ to be compared with 
still have large uncertainties and the results from the existing 
theoretical models of $X(3872)$ are not yet conclusive, more 
experimental and theoretical studies of $X(3872)$ will be needed. 

\vspace{0mm}
\section*{Acknowledgments} \vspace{-0mm}
The author would like to thank Prof. B.~D.~Yabslay and Prof. Y.~Sakai, 
the Belle Collaboration, and Prof. Onogi, YITP for discussions. 
This work was motivated by the discussions with them. 


\end{document}